\documentclass[aps,showpacs,preprintnumbers,superscriptaddress,nofootinbib,preprint]{revtex4}
\usepackage{amssymb}
\usepackage{amsmath}
\usepackage{graphicx}
\usepackage{dcolumn}
\usepackage{bm}

\begin{document}

\title{Diffractive dissociation including pomeron loops in zero transverse
dimensions}
\author{Arif~I.~Shoshi}
\email{shoshi@physik.uni-bielefeld.de}
\affiliation{Fakult{\"a}t f{\"u}r Physik, Universit{\"a}t Bielefeld, D-33501 Bielefeld,
Germany}
\author{Bo-Wen Xiao}
\email{bowen@phys.columbia.edu}
\affiliation{Department of Physics, Columbia University, New York, NY, 10027, USA}

\begin{abstract}
We have recently studied the QCD pomeron loop evolution equations in zero
transverse dimensions~\cite{Shoshi:2005pf}. Using the techniques developed
in~\cite{Shoshi:2005pf} together with the AGK cutting rules, we present a
calculation of single, double and central diffractive cross sections (for
large diffractive masses and large rapidity gaps) in zero transverse
dimensions in which all dominant pomeron loop graphs are consistently
summed. We find that the diffractive cross sections unitarise at large
energies and that they are suppressed by powers of $\alpha_s$. Our
calculation is expected to expose some of the diffractive physics in
hadron-hadron collisions at high energy.
\end{abstract}

\date{\today }
\preprint{BI-TP 2006/20}
\preprint{CU-TP-1148}
\pacs{13.85-t;13.85.Hd; 11.55.-m ;11.55.Bq}
\maketitle

\newpage

\section{Introduction}

QCD evolution equations have been recently established which include pomeron
loops (or fluctuations)~\cite{Mueller:2005ut+X}. The main feature of these
equations, as compared to the Balitsky-JIMWLK~\cite%
{Balitsky:1995ub+X,Jalilian-Marian:1997jx+X} or Kovchegov equations~\cite%
{Kovchegov:1999yj+X}, are the violation of the geometrical scaling of the $T$%
-matrix and a modified energy dependence for the saturation momentum~\cite%
{Mueller:2004se+X}. Recently the effect of fluctuations also on
single diffractive dissociation~\cite{Hatta:2006hs} has been
studied and compared with previously obtained mean-field-like
results~\cite{Munier:2003zb} at high energies.

In a recent paper~\cite{Shoshi:2005pf} we have studied the QCD
pomeron loop equations in zero transverse dimensions in which case
one can do the calculation of any pomeron loop graph analytically
(see
also~\cite{Mueller:1996te,Rembiesa:2005gj,Kozlov:2006zj+X,Kozlov:2006cu,Blaizot:2006wp,Bondarenko:2006rh,Kovner:2005qj,Iancu:2004iy}).
The pomeron loop equations are a hierarchy of equations which in
zero transverse dimensions take the form
\begin{eqnarray}
\frac{dn^{(k)}}{dy}=k\alpha n^{(k)}+k(k-1)\alpha n^{(k-1)}-k\beta n^{(k+1)}
\ ,  \label{h1}
\end{eqnarray}
where $n^{(k)}$ is a normal ordered number operator defined as $\left\langle
\widetilde{n}(\widetilde{n}-1)\cdots (\widetilde{n} -k+1)\right\rangle$
which represents the expectation value of $k$-pomerons during the evolution.
The single terms in eq.~(\ref{h1}) have the following physical meaning: $%
k\alpha n^{(k)}$ is the BFKL growth term ($\alpha$ here corresponds to $%
\alpha _{P}-1=\frac{ 4\alpha _{s}N_{c}}{\pi }\ln 2$ in the real BFKL
equation), $k(k-1)\alpha n^{(k-1)}$ describes fluctuations (pomeron
splittings) and $k\beta n^{(k+1)}$ recombinations (pomeron mergings). Eq.~(%
\ref{h1}) is the zero-transverse-dimensional analog of the real QCD equations
(see~\cite{Mueller:2005ut+X}). One can easely show that (\ref{h1}) leads to
frame-independent scattering amplitudes when $\beta=\alpha \alpha_s^2$.

The scattering amplitudes have been calculated in
Ref.~\cite{Shoshi:2005pf} by treating the recombination terms as
small perturbations. In a recent paper together with our
collaborators~\cite{Bondarenko:2006rh} we have shown that at
rapidity $Y \gg 1/\alpha_s^2 \alpha$, the perturbative treatment
of the recombination terms becomes inaccurate and needs to be
replaced by more complete calculations (This limit is also
indicated in our corrections to the LO result given in Eq.~(15) in
Ref.~\cite{Shoshi:2005pf}). Thus, the results for diffractive
cross sections shown in this paper are only reliable up to $Y
\approx 1/\alpha^3_s$.

\begin{figure}[htbp]
\begin{center}
\includegraphics[width=12cm]{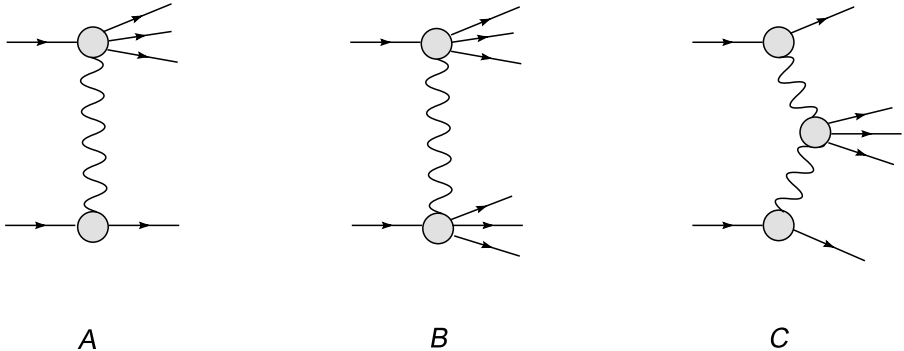}
\end{center}
\caption[*]{Typical diffractive dissociation graphs: Single
diffractive
  scattering (A), double diffractive scattering (B), central
diffractive scattering (C).} \label{d3}
\end{figure}

In this paper we use the techniques which we have developed in~\cite%
{Shoshi:2005pf} together with the AGK cutting
rules~\cite{Abramovsky:1973fm} to calculate single, double and
central diffractive cross sections in zero transverse dimensions
(see Fig.~\ref{d3}). We do consider only the case where the
diffractive masses, rapidity gaps and the total rapidity are
large. All the relevant pomeron graphs are included in the
calculations. In four dimensional QCD, the analogous calculations
are not yet available. So far only the single diffractive
scattering in real QCD including pomeron loops has been
calculated, however, only in the kinematical region where the
diffractive mass is not too large~\cite{Hatta:2006hs}. Thus, our
calculation in zero transverse dimensions may give indications
about the diffractive mass dependence of the diffractive cross
section in four dimensional QCD at very large diffractive masses.
We expect the energy dependence of the QCD high energy diffractive
scattering would bear some resemblances of our result.

A longstanding problem has been accompanying diffractive dissociation: The
lowest order pomeron contribution to diffractive dissociation disagrees with
the Fermilab data (see e.g.,~\cite{Goulianos:1995wy}) and it violates
unitarity at very high energies. There have been several attempts to solve
this problem by considering the renormalization of the pomeron flux~\cite%
{Goulianos:1995wy}, a larger impact parameter~\cite{Mueller:1997ik+X} or a
smaller survival probability~\cite{Gotsman:1998mm} at high energies. In this
work we show that single, double and central diffractive cross sections would
naturally fulfill unitarity limits when multiple pomeron exchanges are taken
into account in addition to the lowest order pomeron exchange.

Furthermore we do find that the differential diffractive cross sections
(differential with respect to the masses) are suppressed by powers of $%
\alpha_s$. We will show that this suppression is characteristic for
differential diffractive cross sections and emerges from pomeron
splittings/mergings in the $t$-channel which are needed in order to
calculate the differential diffractive cross sections.

The paper is organized as follows: We do start with an illustration of the
AGK cutting rules, then go over to the single diffractive cross section,
which is then followed by double diffractive and central diffractive cross
sections.

\section{The AGK cutting rules}

The AGK cutting rules~\cite{Abramovsky:1973fm} (see also~\cite
{Koplik:1975ni+X,Mueller:1996te}) represent the generalization of the optical
theorem for the case of multiple pomeron exchange.  Although AGK cutting rules
have not been proven for QCD, they have been discussed and used in several
publications (see Refs.~\cite{Bartels:2005wa,Kovchegov:1999ji,Treleani:1994at})
. We will show below how the AGK cutting rules lead to expectable results for
particle-particle and particle-nucleus results also within this QCD model, and
then use them to also calculate diffractive scattering processes.

\begin{figure}[tbp]
\begin{center}
\par
\includegraphics[width=13cm]{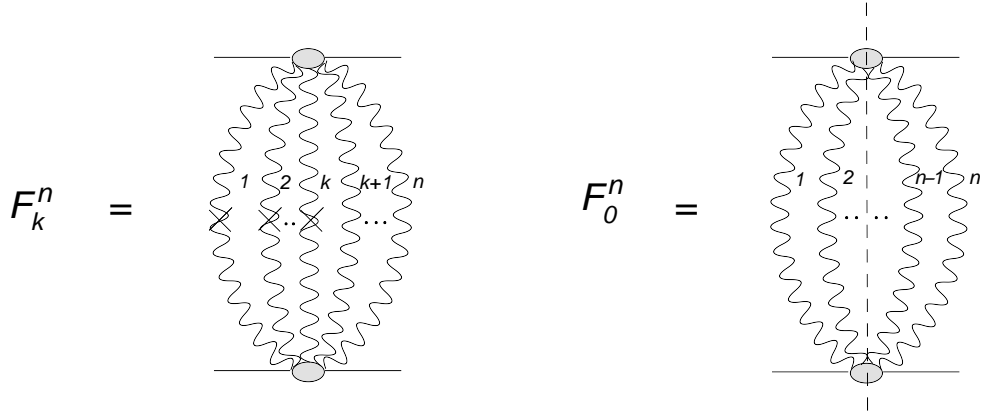}
\end{center}
\caption[*]{$n$-pomeron exchange graphs: In graph A $k$ of the $n$-pomerons
are cut (inelastic cut) and in graph B none of the pomerons is cut (elastic
cut). }
\label{elastic}
\end{figure}

To illustrate the AGK cutting rules, let us consider for instance the total
cross section for particle-particle scattering,
\begin{eqnarray}
\sigma_{tot}=2\sum_{n=1}^{\infty} \left( -1 \right)^{n-1}\ F^{n},
\label{sig_tot}
\end{eqnarray}
where $F^n$ is the amplitude for the exchange of $n$ pomerons. According to
the AGK cutting rules, $\sigma_{tot}$ is related to the elastic cross
section $\sigma_0$ and inelastic cross section $\sigma_{in}$,
\begin{eqnarray}
\sigma_{tot} = \sigma_0 + \sigma_{in} \ ,  \label{AGK_rel}
\end{eqnarray}
which are given by
\begin{eqnarray}
\sigma_0 = \sum_{n=2}^{\infty}\ F^n_0  \label{sig_diff}
\end{eqnarray}
and\footnote{%
In eq.~(\ref{sig_diff}) $n$ starts from $2$ because there have to be at
least two pomerons, one pomeron on either side of the elastic cut. One  side
of the cut can be viewed as initial state of the scattering while the  other
side can be understood as the final state.}
\begin{eqnarray}
\sigma_{in} = \sum_{k=1}^{\infty}\ \sigma_k = \sum_{k=1}^{\infty}\
\sum_{n=k}^{\infty }F_{k}^{n} \ .  \label{sig_inel}
\end{eqnarray}
In the above equations $F^n_0$ represents the $n$-pomeron exchange graph
where no pomeron is cut (elastic cut) as shown in Fig.~\ref{elastic}B and $%
F^n_k$ is the $n$-pomeron exchange graph with $k$ pomerons being
simultaneously cut (inelastic cut) as shown in Fig.~\ref{elastic}A (cut
pomerons are marked by crosses). In the next sections we will introduce also
the diffractive cut as shown for example in Fig.~\ref{sd}A where only a part
of a pomeron is cut. The diffractive cut generates diffractive
(dissociation) cross sections while the inelastic cut (cut of an entire
pomeron) leads to a uniform production of final state particles in rapidity.

For convenience, let us summarize the AGK cutting rules, which allow us to
calculate $F^n_0$ and $F^n_k$ and the diffractive cross sections in the next
sections:

\begin{itemize}
\item No matter how many pomerons are cut, there is always one
and only one cut which indicates the separation between the initial and
final states of the scattering;

\item Each cut-pomeron gives an extra factor of $\left(
-2\right)$ , which can be understood as a result of the discontinuity of the
pomeron amplitude;

\item Each un-cut pomeron obtains an extra factor of $2$ since
it can be placed on either side of the cut.
\end{itemize}


Thus, applying the above AGK cutting rules on Fig.~\ref{elastic}, we obtain
\begin{eqnarray}
F^n_0 =(-1)^n\,2\,(2^{n-1}-1)\,F^n  \label{Fn_0}
\end{eqnarray}
and
\begin{eqnarray}
F_{k}^{n}=\left( -1\right) ^{n}\left( -2\right) ^{k}\left( 2\right)
^{n-k}C_{n}^{k}F^n  \label{Fn_k}
\end{eqnarray}
where $C_n^k = n!/[k!\,(n-k)!]$ is the number of selections of $k$ pomerons
out of $n$ pomerons. It is easy to verify Eq.~(\ref{AGK_rel}) by
substituting the above expressions for $F^n_0$ and $F^n_k$ in Eq.~(\ref%
{sig_diff}) and Eq.~(\ref{sig_inel}), respectively.

To gain confidence in the AGK cutting rules, let us show that the
cross sections obtained using the AGK cutting rules and the
pomeron calculus do agree with known unitarity limits at high
energies. For particle-particle scattering the amplitude for
$n$-pomeron exchanges at leading order~\cite{Shoshi:2005pf} is
$F^n=n!\left( \alpha _{s}^{2}e^{\alpha Y}\right) ^{n}$ (the factor
$(-1)^n$ has already been taken into account in
Eqs.(\ref{sig_tot},\ref{sig_diff},\ref{sig_inel})). Inserting this
amplitude in Eqs.(\ref{sig_tot},\ref{sig_diff},\ref{sig_inel}),
one finds $\sigma _{tot}\simeq 2$, $ \sigma _{inel}\simeq 1$ and
$\sigma _{0}\simeq 1$ in large $Y$ limit, which is in agreement
with the black disk limit. For particle-nucleus scattering, now
with $F^n=\left( \alpha _{s}^{2}e^{\alpha Y}L\right)^{n}$, where
$L$ is the number of hadrons in the nucleus and $\alpha^2 L \gg
1$, it is straightforward to obtain the following results for the
total, elastic (diffractive) and inelastic cross sections,
\begin{eqnarray}
\sigma _{tot} &=&\frac{2\alpha _{s}^{2}e^{\alpha Y}L}{1+\alpha
_{s}^{2}e^{\alpha Y}L}, \\
\sigma _{0} &=&\frac{2\left( \alpha _{s}^{2}e^{\alpha Y}L\right) ^{2}}{%
\left( 1+\alpha _{s}^{2}e^{\alpha Y}L\right) \left( 1+2\alpha
_{s}^{2}e^{\alpha Y}L\right) }, \\
\sigma _{in} &=&\frac{2\alpha _{s}^{2}e^{\alpha Y}L}{1+2\alpha
_{s}^{2}e^{\alpha Y}L} ,
\end{eqnarray}
which do as well respect unitarity limits at very high rapidity.

\section{Single diffractive scattering}

\label{sec:1}
\begin{figure}[tbp]
\begin{center}
\includegraphics[width=10cm]{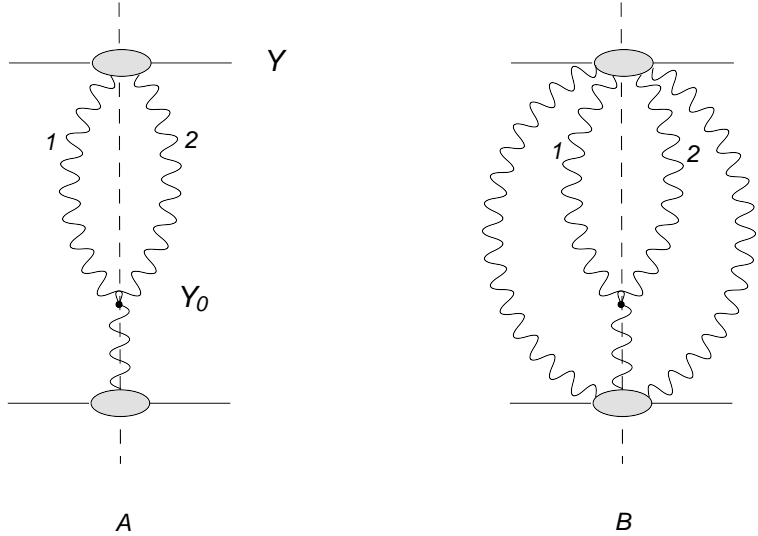}
\end{center}
\caption[*]{Diffractive cut of the multiple pomeron exchange graphs which
generate single diffractive dissociation: Diagram A is the lowest order
graph and diagram B is a higher order graph.}
\label{sd}
\end{figure}

The single diffractive process, $p + p \rightarrow p+X$, as sketched in Fig.~%
\ref{d3}, is a process where one of the $p$-particles breaks up into a
"diffractive state" $X$ which is separated by a rapidity gap from the $p$%
-particle in the final state which remains intact. In the language
of pomerons, the single diffractive production is generated by a
diffractive cut of a multiple pomeron exchange graph as shown in
Fig.~\ref{sd}A (lowest order graph) and Fig.~\ref{sd}B (higher
order graph). The appropriate variables to describe the single
diffractive scattering process are the rapidities $Y = \ln
\frac{s}{m_p^2}$ and $Y_0 = \ln \frac{M_X^2}{m_p^2}$, where $s$ is
the square of the center of mass energy, $M_{X}$ is the
diffractive mass, and $m_{p}$ is the rest mass of the particle
$p$. In this picture, the rapidity gap $Y-Y_0$ is easily
understood since there is no cut through any of the pomerons in
this rapidity window.

We calculate the single diffractive cross section for the case where $Y$, $%
Y_{0}$ and $Y-Y_{0}$ are large. In this kinematical window the dominant
graphs are those shown in Fig.~\ref{sd}. The leading order contribution of
the dominant graphs is calculated in this work. (The subdominant graphs which we
neglect are those where pomeron loops are formed in the diffractive region $%
0 < y < Y_0$ or in the rapidity gap region $Y_0 < y < Y$. The loops
in these windows are smaller as compared to the dominant loops stretching over
the whole rapidity $Y$ and give therefore small corrections, proportional to $\exp\left( -\alpha Y_{0}\right)$ or $\exp%
\left[-\alpha \left(  Y-Y_{0}\right)\right]$, in comparison to the main result
coming from the large loops. In addition there are
also other subleading contributions neglected here which are $\alpha_s$
suppressed as compared to the main result which are discussed in detail in Ref.~\cite{Shoshi:2005pf}.).

According to the topology of Fig.~\ref{sd} and the AGK cutting rules
summarized above, we get for the differential single diffractive cross
section \footnote{%
We have $d\sigma_{SD}/dY_0$  instead of $\sigma_{SD}$ in Eq.~(\ref{single})
since we have considered the production of a particular diffractive mass $M_X
$ (or $Y_0  =\ln(M_X^2/m_p^2)$).}
\begin{eqnarray}
\frac{d\sigma _{SD}\left( Y,Y_{0}\right) }{dY_{0}}=\sum_{n=2}^{\infty
}\left( -1\right) ^{n}2^{n-1}F^n\left( Y,Y_{0}\right) ,  \label{single}
\end{eqnarray}%
where the factor $2^{n-1}$ indicates the number of different diagrams when
the diffractive cut is made and $F^{n}$ is the uncut single diffractive
amplitude for the case of $n$-pomeron exchange.

We extract $F^{n}$ from the $T$-matrix for the uncut graphs. The latter is
calculated within the zero transverse dimensional model~\cite{Shoshi:2005pf}
which we have recently developed. In the limit of $\exp \left( \alpha
Y_{0}\right) \gg 1$ and $\exp \left[ \alpha \left( Y-Y_{0}\right) \right]
\gg 1$, the dominant term is the one where $k$ pomeron splittings occur at $%
Y=0$,
\begin{equation*}
n_{k}^{\left( k+1\right) }(Y)\simeq \left( k+1\right) !e^{\left( k+1\right)
\alpha Y}\ ,
\end{equation*}%
the last pomeron splitting occurs at $Y_{0}$,
\begin{eqnarray}
n_{k}^{(k+2)}(Y,Y_{0}) &\simeq &\left( k+2\right) \left( k+1\right)
\int_{0}^{Y}\alpha dyn_{k}^{\left( k+1\right) }(y)e^{(k+2)\alpha \left(
Y-y\right) }\delta \left( y-Y_{0}\right)   \label{k2} \\
&\simeq &\alpha \left( k+1\right) \left( k+2\right) !e^{\left( k+2\right)
\alpha Y-\alpha Y_{0}}\ ,
\end{eqnarray}%
and all the $k+1$ pomeron mergings occur at $Y$
\begin{equation*}
n_{k}^{(1)}(Y,Y_{0})\simeq \left( -1\right) ^{k+1}\alpha \left( k+1\right)
\left( k+2\right) !x^{k}e^{2\alpha Y-\alpha Y_{0}}\ ,
\end{equation*}%
where $x=\alpha _{s}^{2}e^{\alpha Y}$ and $T_{k}^{(1)}(Y,Y_{0})=\alpha
_{s}^{2}\ n_{k}^{(1)}(Y,Y_{0})$. Now, from $T_{k}^{(1)}(Y,Y_{0})=\left(
-1\right) ^{k}\alpha \left( k+1\right) \left( k+2\right) !x^{k+2}e^{-\alpha
Y_{0}}$, we extract $F^{n}=\alpha \left( n-1\right) n!x^{n}e^{-\alpha Y_{0}}$
by using $F^{n}=(-1)^{n}T_{n-2}^{(1)}$, which inserted into the diffractive
cross section given in Eq.~(\ref{single}) yields
\begin{eqnarray}
\frac{d\sigma _{SD}\left( Y,Y_{0}\right) }{dY_{0}} &\simeq &\alpha
e^{-\alpha Y_{0}}\sum_{n=2}^{\infty }\left( -1\right) ^{n}2^{n-1}\left(
n-1\right) n!x^{n},  \label{sdsum} \\
&=&-\frac{1}{2}\alpha e^{-\alpha Y_{0}}z^{2}\frac{d^{2}}{dz^{2}}\left[
\Gamma \left( 0,\frac{1}{z}\right) e^{\frac{1}{z}}\right] ,  \label{sdsum1}
\end{eqnarray}%
where we have used the definition $z=2\alpha _{s}^{2}e^{\alpha Y}$. The
first term of the sum in Eq.~(\ref{sdsum}) reproduces the energy dependence
of the first order Regge prediction for single diffractive processes, and
other terms in the summation are contributions of higher order dominant
graphs. In the high energy limit, when the diffractive mass (or $Y_{0}$) and
the rapidity gap ($Y-Y_{0}$) are kept large, we find
\begin{equation}
\frac{d\sigma _{SD}\left( Y,Y_{0}\right) }{dY_{0}}\simeq
\frac{1}{2}\alpha e^{-\alpha Y_{0}}\left[ 1+\frac{1+2\gamma
_{E}-2\ln \left( 2\alpha _{s}^{2}e^{\alpha Y}\right) }{2\alpha
_{s}^{2}e^{\alpha Y}}\right] \ ,\label{sd_y_asy}
\end{equation}%
or, in terms of the diffractive mass,
\begin{equation}
M_{x}^{2}\frac{d\sigma _{SD}\left( s,M_{x}^{2}\right) }{dM_{x}^{2}}\simeq
\frac{1}{2}\alpha \left( \frac{m_{p}^{2}}{M_{x}^{2}}\right) ^{\alpha }\ .
\end{equation}%
where $\gamma _{E}=0.577$ is the Euler constant.

Here, we would like to comment that Eq.~(\ref{sd_y_asy}) is approximately true
when $\alpha Y_0$ and $\alpha (Y-Y_0)$ are relatively large, where we find the
result $\frac{1}{2}\,\alpha\, e^{- \alpha Y_0}$. The expectation that all
inelastic diffraction vanishes for $Y \to \infty$~\cite{Hatta:2006hs} can not be
shown in our model since our model does not apply for $Y \gg 1/\alpha_s^3$.  The
origin of our result is obvious: There are ($k+1$) pomerons in the region $0 <
y < Y_0 $. At $Y_0$ another pomeron is produced through pomeron splitting
enforced by the delta-function in Eq.~(\ref{k2}), see also Fig.~\ref{sd}, adding
a contribution $\alpha e^{\alpha (Y-Y_0)}$ to the ($k+1$) pomerons of length $Y
$. This explains the appearance of $\alpha e^{-\alpha Y_0}$. The remaining $%
e^{\alpha Y}$-dependence together with the other pomerons of the same length
$Y$ do give a sum over all pomeron exchanges which naturally unitarizes the
single diffractive cross section at large $Y$. The factor of $1/2$ in (%
\ref{sd_y_asy}) is a combinatorial factor emerging after the diffractive cut
is made.

Another phenomenon not observed so far elsewhere is that the single
diffractive cross section in (\ref{sd_y_asy}) is suppressed by a factor of $%
\alpha$. This factor appears through the pomeron splitting at $Y_0$. A
suppression by powers of $\alpha$ is characteristic for differential
diffractive cross sections because pomeron splittings in the $t$-channel
(see Figs.~\ref{sd},\ref{dd},\ref{cd}) are needed in order to calculate
them. In the next sections we show how also double and central diffractive
cross sections are suppressed by powers of $\alpha$.

It is tempting to integrate Eq.~(\ref{sd_y_asy}) over $Y_0$ from $0$ to $Y$
to obtain a formula for $\sigma_{SD}(Y)$. However, this wouldn't be right
since Eq.~(\ref{sd_y_asy}) is not valid for small values of diffractive
masses and rapidity gaps. When diffractive masses or rapidity gaps are
small, many other graphs have to be considered besides the dominant graphs
shown in Fig.~\ref{sd}. In this case the resummation of all order graphs is
no longer under control within our formalism.

In Ref.~\cite{Hatta:2006hs} the (differential) single diffractive cross
section in four dimensions including pomeron loops has been calculated.
However, the calculation is only applicable in the region where the
diffractive mass $Y_0$ is small, $Y_0 \ll \pi/(\alpha_s
N_c)\,\ln(\pi^2/\alpha_s^2)$. On the other hand, our result applies when the
diffractive mass is large, $\exp[\alpha Y_0] \gg 1$. It would be interesting
to see whether our result remains valid also in four dimensional QCD.

\section{Double diffractive scattering}

\label{sec:4}
\begin{figure}[tbp]
\begin{center}
\includegraphics[width=12cm]{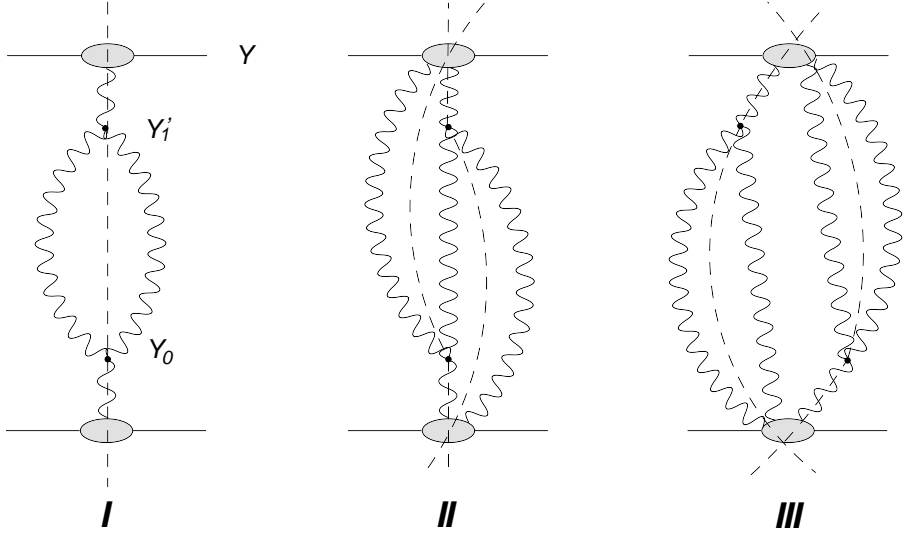}
\end{center}
\caption[*]{Three topologically different types of the diffractively cut
multiple pomeron exchange graphs which generate double diffractive
dissociation. Diagram I (type I) is the lowest order graph; Diagram II (type
II) is a higher order graph which requires at least 3 pomerons exchanged;
Diagram III (type III) is another higher order graph which needs at least 4
pomerons exchanged.}
\label{dd}
\end{figure}
The double diffractive scattering process, $p + p \rightarrow X + X^{\prime}$
(see Fig.~\ref{d3}B), is generated by the three topologically different
types of diffractively cut multiple pomeron exchanges as shown in the lowest
order in Fig.~\ref{dd}. We use below the variables $Y_{0}=\ln \frac{M_{X}^{2}%
}{m_{p}^{2}}$ and $Y_{1}=Y-Y^{\prime }_{1}=\ln \frac{M_{X}^{\prime 2}}{%
m_{p}^{2}}$, where $M_{X}$ and $M_{X}^{\prime}$ are the diffractive masses,
to express the double diffractive cross section. The rapidity gap between
the diffractive states $X$ and $X^{\prime}$ is $Y-Y_0 -Y_1$.

Let us first compute the dominant contribution of the un-cut amplitude for
the case of multiple pomeron exchange with one pomeron splitting at $Y_{0}$
and one pomeron merging at $Y_{1}$. Further we consider the case where $Y_0$
and $Y_1$ are kept large. The result after the last pomeron splitting at $Y_0
$ which is given by Eq.~(\ref{k2}), when followed by a pomeron merging at $%
Y_{1}^{\prime}$, gives
\begin{eqnarray}
n_{k}^{(k+1)}(Y_{1}^{\prime },Y_{0},Y) &\simeq &-\alpha _{s}^{2}\left(
k+1\right) \int_{0}^{Y}\alpha dyn_{k}^{\left( k+2\right) }(y)e^{(k+1)\alpha
\left( Y-y\right) }\delta \left( y-Y_{1}^{\prime }\right) , \\
&=&\left( -\alpha _{s}^{2}\right) \left( k+2\right) !\left( k+1\right)
^{2}\alpha ^{2}e^{\left( k+1\right) \alpha Y+\alpha Y_{1}^{\prime }-\alpha
Y_{0}} \ ,
\end{eqnarray}
and after $k$ more pomeron mergings at rapidity $Y$ (see~\cite{Shoshi:2005pf}%
), one ends up with $T_{k}^{(1)}(Y,Y_{1},Y_{0})\simeq \left( -1\right)
^{k+1}\alpha ^{2}\left( k+1\right) ^{2}\left( k+2\right) !x^{k+2}e^{-\alpha
Y_{1}-\alpha Y_{0}}$, where $Y_{1}=Y-Y_{1}^{\prime }$.

The resulting $F_{DD}^{n}=\alpha ^{2}\left( n-1\right) ^{2}n!x^{n}e^{-\alpha
Y_{0}-\alpha Y_{1}}$ has the following interpretation: There are $(n-1)^{2}$
different graphs in the $n$-pomeron exchange case, each of them contributing
by the same amplitude $\alpha ^{2}\,n!\,x^{n}\,e^{-\alpha Y_{0}-\alpha Y_{1}}
$. Among these $\left( n-1\right) ^{2}$ pomeron exchange graphs, there are
three topologically different types when the diffractive cut is made (e.g.,
see Fig.~\ref{dd}). In the $n$-pomeron exchange case, it is easy to find
that there are $N_{I}^{\left( n\right) }=\left( n-1\right) $ diagrams of the
first type, $N_{II}^{\left( n\right) }=2\left( n-2\right) $ diagrams of the
second type and $N_{III}^{\left( n\right) }=\left( n-2\right) \left(
n-3\right) $ diagrams of the third type. (Evidently, $N_{I}^{\left( n\right)
}+N_{II}^{\left( n\right) }+N_{III}^{\left( n\right) }=\left( n-1\right) ^{2}
$.) When the cut is made, the number of diagrams of type I becomes $%
N_{I,cut}^{\left( n\right) }=\left( n-1\right) 2^{n-1}$, the number of
diagrams of type II turns out to be $N_{II,cut}^{\left( n\right) }=\left(
n-2\right) 2^{n-2}$ and the number of diagrams of type III is $%
N_{III,cut}^{\left( n\right) }=\left( n-2\right) \left( n-3\right) 2^{n-3}$.
With the total number of diagrams after the cut, $N_{cut}^{\left( n\right) }=%
\left[ 4\left( n-1\right) +\left( n-1\right) \left( n-2\right) \right]
2^{n-3}$, we get for the double diffractive cross section
\begin{equation*}
\frac{d\sigma _{DD}\left( Y,Y_{0},Y_{1}\right) }{dY_{0}dY_{1}}=\alpha
^{2}e^{-\alpha Y_{0}-\alpha Y_{1}}\sum_{n=2}^{\infty }\left( -1\right)
^{n}n!x^{n}N_{cut}^{\left( n\right) }\ .
\end{equation*}%
In high energy limit, while $Y_{0}$ and $Y_{1}$ large, one obtains
\begin{equation*}
\frac{d\sigma _{DD}\left( Y,Y_{0},Y_{1}\right) }{dY_{0}dY_{1}}\simeq \frac{1%
}{4}\alpha ^{2}e^{-\alpha Y_{0}-\alpha Y_{1}}\ ,
\end{equation*}%
or, equivalently,
\begin{equation*}
M_{x}^{2}M_{x}^{\prime 2}\frac{d\sigma _{DD}\left( s,M_{x}^{2},M_{x}^{\prime
2}\right) }{dM_{x}^{2}dM_{x}^{\prime 2}}\simeq \frac{1}{4}\alpha ^{2}\left(
\frac{m_{p}^{2}}{M_{x}^{2}}\right) ^{\alpha }\left( \frac{m_{p}^{2}}{%
M_{x}^{\prime 2}}\right) ^{\alpha }\ .
\end{equation*}%
The double diffractive cross section equals the product of two
single diffractive cross sections at very large $Y$, which is in
line with the naive expectation $d\sigma _{DD}/dY_{0}dY_{1}\sim
\lbrack d\sigma _{SD}/dY_{0}\ d\sigma _{SD}/dY_{1}]/\sigma _{tot}$
since the double diffractive process can be viewed as two separate
and independent single diffractive processes occurring at both,
the projectile and the target. Therefore one obtains a stronger
$\alpha $ and diffractive mass suppression for the differential
double diffractive cross section as compared with the differential
single diffractive cross section. Also the double diffractive
cross section unitarizes due to the multiple pomeron exchanges,
instead of showing a rapid growth, when $Y$ becomes very large large.

\section{Central diffractive scattering}

\begin{figure}[tbph]
\begin{center}
\includegraphics[width=4cm]{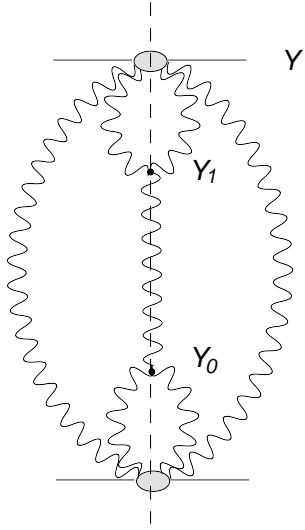}
\end{center}
\caption[*]{Diffractive cut of general multi-pomeron exchange graphs which
generate central diffractive dissociation.}
\label{cd}
\end{figure}
The central diffractive scattering, $p+p\rightarrow p+X+p$ (see Fig.~\ref{d3}%
C), as shown in Fig.~\ref{cd} (higher order graph), is obtained by following
the same procedure as in the previous sections. The un-cut amplitude of $k$%
-th order (i.e., $k$ pomeron exchanges between $0$ and $Y$ in addition to
the lowest order diagram) reads
\begin{equation*}
T_{k}^{(1)}(Y,Y_{1},Y_{0})\simeq \left( -1\right) ^{k+2}2\alpha ^{2}\alpha
_{s}^{2}\left( k+1\right) \left( k+2\right) !x^{k+2}e^{-\alpha Y_{1}+\alpha
Y_{0}}.
\end{equation*}%
When a diffractive cut is made, the diffractive cross section becomes%
\begin{equation*}
\frac{d\sigma _{CD}\left( Y,Y_{0},Y_{1}\right) }{dY_{0}dY_{1}}\simeq \alpha
^{2}\alpha _{s}^{2}e^{-\alpha \left( Y_{1}-Y_{0}\right) }\sum_{n=2}^{\infty
}\left( -1\right) ^{n}\left( n-1\right) n!x^{n}2^{n},
\end{equation*}%
and for high energy scattering but $Y_{1}-Y_{0}=\delta Y=\ln \left( \frac{%
M_{X}^{2}}{m_{p}^{2}}\right) $ large ($M_{X}$ is the central diffractive
mass) it reduces to
\begin{equation}
\frac{d\sigma _{CD}\left( Y,Y_{0},Y_{1}\right)
}{dY_{0}dY_{1}}\simeq \alpha ^{2}\alpha _{s}^{2}e^{-\alpha \delta
Y}=\alpha ^{2}\alpha _{s}^{2}\left(
\frac{m_{p}^{2}}{M_{x}^{2}}\right) ^{\alpha }\ .\label{cd_res}
\end{equation}%
The suppression $\alpha ^{2}\alpha _{s}^{2}$ can be easily understood by
looking at the lowest order graph (Fig.~\ref{cd} without the two pomerons on
both sides of length $Y$): The merging at $Y_{0}$ gives an $\beta $ and the
subsequent splitting at $Y_{1}$ a $\alpha $. The rapidity dependence is $%
\exp [\alpha Y]\exp [\alpha (Y-(Y_{1}-Y_{0}))]$. The remaining factor of $%
\alpha _{s}^{4}$ comes from the coupling of the pomerons to the scattering
particles, thus, giving altogether $\alpha ^{2}\alpha _{s}^{2}\ \exp
[-\alpha (Y_{1}-Y_{0})]\ x^{2}$, where $x=\alpha _{s}^{2}\exp [\alpha Y]$,
which explains the result in (\ref{cd_res}). The higher order graphs
obtained by adding further pomeron exchanges as shown in Fig.~\ref{cd} lead
to a $Y$-dependence in accordance with unitarity limits. The differential
central diffractive cross section is strongly $\alpha _{s}$-suppressed ($%
\alpha \sim \alpha _{s}$) as compared with the differential single (double)
diffractive cross section.

In summary, we systematically calculate single diffractive, double
diffractive and central diffractive cross sections within the zero
transverse dimensional Regge model~\cite{Shoshi:2005pf}. The resummation of
all higher order dominant diagrams shows that diffractive cross sections
gradually approach unitarity limits instead of increasing rapidly at
very large energies. Moreover, we do find powers of $\alpha$ suppression in
the inclusive diffractive cross sections.

\textbf{Acknowledgements}: We acknowledge numerous stimulating discussions
with Alfred Mueller. A. Sh. acknowledges financial support by the Deutsche
Forschungsgemeinschaft under contract Sh 92/2-1 and wishes to thank the
theory group at DESY for hospitality during his visit when this work was
being completed.

\end{document}